\documentclass[iop,apj,numberedappendix,appendixfloats]{emulateapj}

\usepackage{natbib}
\usepackage{longtable}
\usepackage{rotating}
\usepackage[usenames,dvipsnames]{color}
\usepackage{enumerate}
\usepackage{amsmath}
\usepackage{multirow}

\usepackage{ulem}

\usepackage[plainpages=false, colorlinks=true, anchorcolor=blue, linkcolor=blue, citecolor=blue, bookmarks=false]{hyperref}
\usepackage{ulem}

\def\be{\begin{equation}}
\def\ee{\end{equation}}
\newcommand{\bb}{\begin{bmatrix}}
\newcommand{\eb}{\end{bmatrix}}

\shorttitle{A FRB discovered by FAST}
\shortauthors{Zhu et al.}

\begin{document}

\title{A Fast Radio Burst discovered in FAST drift scan survey}

\author{
Weiwei Zhu\altaffilmark{1,6}{*},
Di Li\altaffilmark{1}{*},
Rui Luo\altaffilmark{3},
Chenchen Miao\altaffilmark{2,5},
Bing Zhang\altaffilmark{4},
Laura Spitler\altaffilmark{6},
Duncan Lorimer\altaffilmark{7,8},
Michael Kramer\altaffilmark{6},
David Champion\altaffilmark{6},
Youling Yue\altaffilmark{1},
Andrew Cameron\altaffilmark{3},
Marilyn Cruces\altaffilmark{6},
Ran Duan\altaffilmark{5},
Yi Feng\altaffilmark{2,5,3},
Jun Han\altaffilmark{5,13},
George Hobbs\altaffilmark{3},
Chenhui Niu\altaffilmark{1},
Jiarui Niu\altaffilmark{2,5},
Zhichen Pan\altaffilmark{1},
Lei Qian\altaffilmark{1},
Dai Shi\altaffilmark{3},
Ningyu Tang\altaffilmark{1},
Pei Wang\altaffilmark{1},
Hongfeng Wang\altaffilmark{9,2,10,11},
Mao Yuan\altaffilmark{2,5},
Lei Zhang\altaffilmark{2,5,3},
Xinxin Zhang\altaffilmark{5},
Shuyun Cao\altaffilmark{1},
Li Feng\altaffilmark{1},
Hengqian Gan\altaffilmark{1},
Long Gao\altaffilmark{1},
Xuedong Gu\altaffilmark{1},
Minglei Guo\altaffilmark{1},
Qiaoli Hao\altaffilmark{1},
Lin Huang\altaffilmark{1},
Menglin Huang\altaffilmark{1},
Peng Jiang\altaffilmark{1},
Chengjin Jin\altaffilmark{1},
Hui Li\altaffilmark{1},
Qi Li\altaffilmark{1},
Qisheng Li\altaffilmark{1},
Hongfei Liu\altaffilmark{1},
Gaofeng Pan\altaffilmark{1},
Bo Peng\altaffilmark{1},
Hui Qian\altaffilmark{1},
Xiangwei Shi\altaffilmark{1}, 
Jinyuo Song\altaffilmark{1},
Liqiang Song\altaffilmark{1},
Caihong Sun\altaffilmark{1},
Jinghai Sun\altaffilmark{1},
Hong Wang\altaffilmark{1},
Qiming Wang\altaffilmark{1},
Yi Wang\altaffilmark{1},
Xiaoyao Xie\altaffilmark{1},
Jun Yan\altaffilmark{1},
Li Yang\altaffilmark{1},
Shimo Yang\altaffilmark{1},
Rui Yao\altaffilmark{1},
Dongjun Yu\altaffilmark{1},
Jinglong  Yu\altaffilmark{1},
Chengmin Zhang\altaffilmark{1},
Haiyan Zhang\altaffilmark{1},
Shuxin Zhang\altaffilmark{1},
Xiaonian Zheng\altaffilmark{1},
Aiying Zhou\altaffilmark{1},
Boqin Zhu\altaffilmark{1},
Lichun Zhu\altaffilmark{1},
Ming Zhu\altaffilmark{1},
Wenbai Zhu\altaffilmark{1},
Yan Zhu\altaffilmark{1}
}

\affil{
$^1$ CAS Key Laboratory of FAST, NAOC, Chinese Academy of Sciences,
Beijing 100101, China;\\
$^2$ University of Chinese Academy of Sciences, Beijing 100049, China\\
$^3$ CSIRO Astronomy and Space Science, PO Box 76, Epping, NSW 1710, Australia\\
$^4$ Department of Physics and Astronomy, University of Nevada, Las Vegas, Las Vegas, NV 89154, USA\\
$^5$ National Astronomical Observatories, Chinese Academy of Sciences, Beijing 100012, China\\
$^6$ Max-Planck-Institut für Radioastronomie, Auf dem Hügel 69, D-53121 Bonn, Germany\\
$^7$ Department of Physics and Astronomy, West Virginia University, P.O. Box 6315, Morgantown, WV
26506, USA\\
$^8$ Center for Gravitational Waves and Cosmology, Chestnut Ridge Research Building, Morgantown, WV
26505, USA\\
$^{9}$ Image Processing and Pattern Recognition Laboratory, School of Artificial Intelligence, Beijing Normal University,
Beijing 100875, China; \\
$^{10}$ School of Information Management, Dezhou University, Dezhou 253023, China;\\
$^{11}$ Institute for Astronomical Science, Dezhou University, Dezhou 253023, China;\\
$^{12}$ National Astronomical Data Center, Beijing 100101, China;\\
}

\email{{*}Email: zhuww@nao.cas.cn;dili@nao.cas.cn}




\begin{abstract}
We report the discovery of a highly dispersed fast radio burst, FRB~181123, from an analysis of $\sim$1500~hr of drift-scan survey data taken using the Five-hundred-meter Aperture Spherical radio Telescope (FAST). The pulse has three distinct emission components, which vary with frequency across our 1.0--1.5~GHz observing band. We measure the peak flux density to be $>0.065$~Jy and the corresponding fluence $>0.2$~Jy~ms. Based on the observed dispersion measure of 1812~cm$^{-3}$~pc, we infer a redshift of $\sim 1.9$. From this, we estimate the peak luminosity and
isotropic energy to be $\lesssim 2\times10^{43}$~erg~s$^{-1}$ and  $\lesssim 2\times10^{40}$~erg, respectively. With only one FRB from the survey detected so far, our constraints on the event rate are limited. We derive a 95\% confidence lower limit for the event rate of 900 FRBs per day for FRBs with fluences $>0.025$~Jy~ms. 
We performed follow-up observations of the source with FAST for four hours and have not found a repeated burst. We discuss the implications of this discovery for our understanding of the physical mechanisms of FRBs.
\end{abstract}

\section{Introduction}
\label{sec:intro}
Fast Radio Bursts (FRBs) are bright millisecond-duration radio bursts that are cosmological in origin. They were discovered over a decade ago \citep{lbm+07} and have been studied ever since at major radio observatories including Parkes \citep{lbm+07,ksk+12,tsb+13,zhd+19}, Arecibo  \citep{sch+14}, Green Bank \citep{mkl+15}, Molonglo \citep{cfb+17, ffb+18}.
Recently, two new facilities with a wide field of view have been discovering FRBs in large numbers: the Australian Square Kilometre Array Pathfinder; ASKAP \citep{smb+18} and the Canadian Hydrogen Intensity Mapping Experiment; CHIME \citep{chime}.
Most FRBs seemed to be one-off events, while some are repeating \citep{ssh+16}. 
CHIME has been particularly effective at finding repeating FRBs, with 17 published so far \citep{chime_repeater, chime_8}. 
The origin of these FRBs remains a hot topic of debate and speculation \citep{pww+19}.

The Five-hundred-meter Aperture Spherical radio Telescope
(FAST; \citealt{FAST}) is the largest telescope in the world \citep{jpy+19}.
Due to the FAST's superior sensitivity, \citet{lor18} and \citet{zhang18} predicted that it would be able to detect FRBs of significantly higher dispersion measure (DMs) than those from less-sensitive telescopes. Since high-DM sources are most likely very luminous, FAST
surveys could help to constrain the high-end of the FRB luminosity function and enable more cosmological applications from FRBs \citep{zhang18}. As a first step toward this goal,
we report here a highly dispersed FRB from commissioning observations of FAST. In \S 2, we describe the observations and method used in discovering the FRB event,  in \S 3.1, we present the FRB detection, in \S 3.2 we derive the constraint on the FAST FRB event rate, in \S 4. We summarize the results and discuss the implication of this first blind-search FRB discovery.

\section{Observations}
\subsection{FAST Drift scan survey}

The Commensal Radio Astronomy FAST Survey (CRAFTS\footnote{\url{http://crafts.bao.ac.cn}}, \citealt{li2018})  is a multi-purpose all-sky survey designed to obtain data streams for pulsar searching, transients searching, HI imaging and HI galaxies simultaneously. The survey began testing in August 2017, initially using a single-beam wide-band receiver covering 270--1620~MHz. After May 2018, the survey started using the FAST L-band Array of Nineteen Beams (FLAN), which covers 1050--1450~MHz band with a system temperature of about 20~K \citep{li2018}. 
The drift scan survey typically happens at night (Beijing time from 9 pm to 8 am),  during which time no other observations are scheduled. A total of 138 nights of observations were conducted, and $\sim$1500 hours of 19-beam observations were taken from May 2018 to November 2018, when the burst event was discovered. Data taken subsequently are still being processed.
While we report on FRB searches here, we note that the CRAFTS survey has already discovered over 100 new pulsars\footnote{\url{http://crafts.bao.ac.cn/pulsar}} \citep{qpl+19, zlh+19}.

The original FAST data were written in PSRFITS format \citep{hvm04} with two polarizations and 8-bit sampling at 196.608~$\mu$s intervals and with 4096 spectral channels between 1000 and 1500~MHz. Due to the large data volume, we sum the two polarizations and compress the data to 1-bit before further processing. In the following sections, the signal searching and significance calculations are both based on the single bit summed data, and we include a 
degradation factor of 33\% to account for the loss in signal-to-noise during data compression.
The resulting system parameters we adopt are an average system temperature of 23\,K including contributions from cosmic microwave background, the foreground sky, earth atmosphere and radiation from the surrounding terrain and an effective telescope gain of 10\,K/Jy for beam 17 (15\,K/Jy before digitization loss) \citep{jpy+19,jth+20}.

\subsection{Single Pulse Search System}

FRB~181123 was identified by a novel GPU-based single-pulse search system that integrates the PICS AI software \citep{zbm+14} for selecting single-pulse candidates with the FAST multibeam data. This system uses GPUs to dedisperse the original data streams from each beam into eight subbands for 4096 trial DMs in the range 8.7---9211~pc~cm$^{-3}$. The DM step $\Delta$DM is determined by:
\begin{equation}
C \, \Delta \rm DM (\frac{1}{\nu^2_{\rm min}}-\frac{1}{\nu^2_{\rm max}}) = s \sqrt{\tau^2_{\rm samp} + \tau^2_{\rm pulse} + \tau^2_{\rm smear}},
\end{equation}
here the left hand side is the pulse broadening across the whole band due to one DM step, the right hand side is the pulse broadening in the lowest channel composed of sample time $\tau_{\rm samp}=$196.608~$\mu$s, an assumed minimal pulse width $\tau_{\rm pulse}=0.5$~ms, and the inter-channel DM smearing $\tau_{\rm smear}=2C\delta \nu/\nu^3_{\rm min}$, where $\delta \nu=0.122$~MHz is the channel width; Here $C=4148.808$~MHz$^2$pc$^{-1}$cm$^3$s is the dispersion constant, $s=2$ is a manually chosen sparseness parameter, $\nu_{\rm max}=1500$~MHz, and $\nu_{\rm min}=1000$~MHz. 
The dedispersed time series in each subband are downsampled by a small factor (usually 8) in the GPUs to match the expected typical pulse width of $\sim1$~ms. The code outputs the dedispersed time series in each subband for each DM trial to memory. We then search for threshold-crossing burst events in a summed time series combining all subbands in CPUs with multiple levels of possible burst widths. 
While searching for bursts, the code uses multi-level wavelets\footnote{\url{https://github.com/PyWavelets/pywt}} \citep{Lee2019}
to reduce red noise and search for significant bursts that pass a threshold of 7$\sigma$. 
This burst search normally results in thousands of detections in each dataset. The code then takes the detected signal position in time and DM for each candidate and extracts from the dedispersed subband time series a segment of data that contains the burst signals. This segment of data contains eight frequency subbands and time bins chosen such that the data segment contains 32 times the burst width of data. We refer to these segments as dedispersed snapshots of the burst. Despite some exceptions, most pulsar-like bursts are wide-band signals, their snapshots often contain a full or partial vertical line, which is distinguishable from that of narrow-banded radio frequency interference (RFI). We then employ the CNN classifier in PICS \citep{zbm+14} that was trained using the frequency-vs-phase subplot of the {\it PRESTO}\footnote{\url{https://github.com/scottransom/presto}} \citep{2011ascl.soft07017R} candidate plots. 

The image pattern of a real pulse in our snapshots resembles those in the frequency versus phase subplot in pulsar candidates, i.e., a vertical line. Our experiments show that the PICS-CNN classifier was able to rank the most pulsar-like burst snapshot to the top of all snapshots. We pick only the top candidates from these snapshots (usually with a zero-to-one score $>$0.96, as determined by experiments) to form the final output candidate list. These candidates were subsequently plotted and inspected by eye. This GPU single-pulse search system (enabled by PICS) helped in the discovery of over 20 new pulsars in the FAST drift scan survey, including those reported in \citet{qpl+19} and \citet{zlh+19}. This PICS-aided search system uses non-standard ranking criteria. 
Although successful in finding some pulsars, it does not necessarily detect all pulses that cross the event threshold.
A careful study of the recall of this system will be presented in a later contribution (Zhu et al., in prep.) For now, we assume that this system does not find all true signals that cross the threshold. Meanwhile, we also searched the data using the more standard {\it HEIMDALL}\footnote{\url{http://sourceforge.net/projects/heimdall-astro}} \citep{bbbf12} pipeline and will report the results in future publication.

\section{Results}
\label{sec:result}

\subsection{FRB~181123}

FRB~181123 was detected with
a significance of 19$\sigma$ in beam 17 of the multibeam receiver on MJD 58445.
More detailed parameters with uncertainties are summarized in Table 1. 
We searched time series from other beams that are dedispersed with the same DM but found no signal above $3\sigma$ during the same time.
From the logged position of the receiver cabin at the time, we infer that the FRB came from the direction of $l=184^{\degr}$.06, $b=-13^{\degr}$.47 with a positional
uncertainty of 3' based on the full width of the FAST beam at the center frequency of 1250~MHz. The observed DM for this FRB (1812~pc~cm$^{-3}$) is substantially greater than the maximum DM expected from the Galaxy in this direction $\sim 150$~pc~cm$^{-3}$ \citep{ymw16}.

\begin{figure*}
\centering
\includegraphics[scale=0.7]{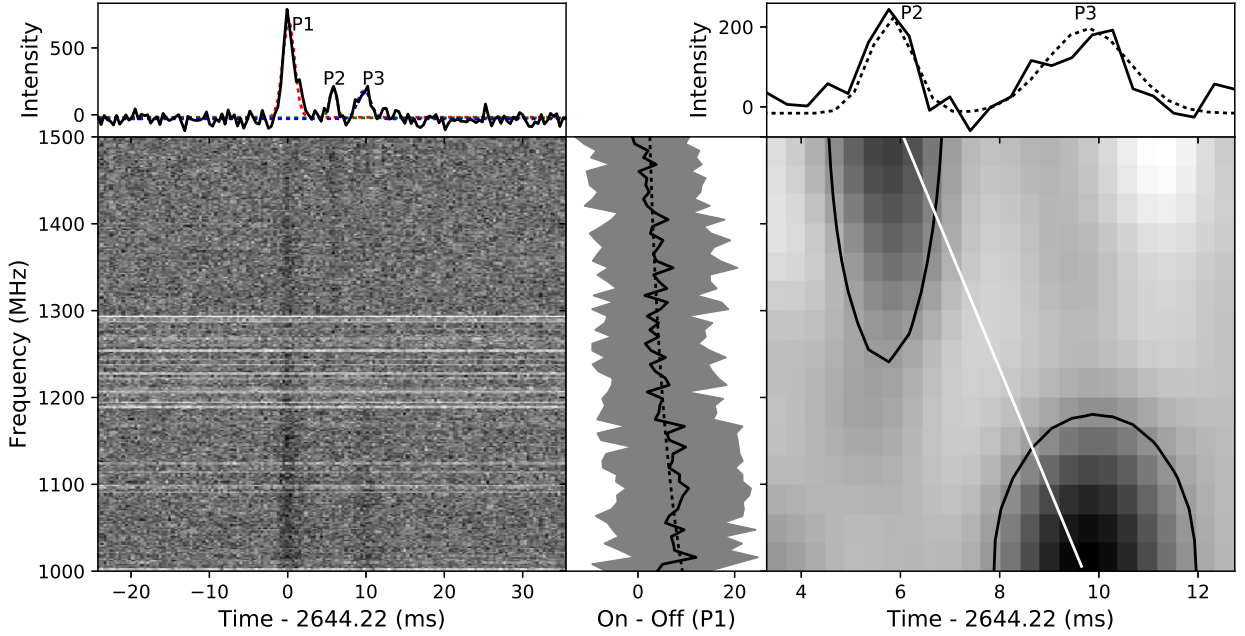} \\ 
\caption{\label{fig:dedisp} 
{\bf Left-Top panel}: The summed pulse profile from the dedispersed pulse, showing two smaller peaks following closely after the main pulse. The red, green, and blue dotted lines show the best Gaussian-fit to the three peaks. The best-fit parameters are listed in Table \ref{tab:par}. {\bf Left panel}: The dedispersed pulse plot showing clear multi-peak structures. The straightness of the pulse indicates a good fit to the $\nu^{-2}$ dispersion law. The horizontal white strips are the results of channels being cleared due to RFI contamination. {\bf Center panel}: The spectrum of FRB 181123 (on-pulse mean spectrum minus the off-pulse mean spectrum for the main peak P1). Here we only take the on-pulse part (24 spectral samples) of P1 for the on-pulse spectrum. We take 400 spectral samples to form the average off-pulse spectrum, 200 on the left of the P1 pulse, and 200 on the right of the P3 pulse.The gray shadow indicates the uncertainty of the on-off spectrum estimated from the root mean squares of the on- and  off-pulse spectra. {\bf Right panel:} The zoomed dynamic spectrum of P2 and P3 smoothed with a Gaussian filter. We fit the two peaks with 2D Gaussian functions to estimate the frequency drift rate between the two peaks. The white line connects the centers of the two best-fit Gaussian functions. {\bf Right-Top panel:} A zoomed view of the pulse profiles of P2 and P3.}
\end{figure*}

Figure~\ref{fig:dedisp} shows a more detailed look at the time--frequency spectrum of FRB~181123, along with the dedispersed pulse profile. The burst shows a multi-peak pulse profile with three distinguishable peaks separated by few milliseconds (labeled as P1, P2, and P3). The measured parameters of these peaks are presented in Table \ref{tab:par}. 
From Gaussian fits to the pulse profiles, we infer that P2 arrives about 5.6~ms after P1, and P3 arrives about 4~ms trailing P2 in the observer's frame, these correspond to 1.9~ms and 1.4~ms delays in the rest frame of the FRB.
Using the radiometer equation to convert our data to a Jansky scale, we measure the specific peak flux to be 65~mJy for P1 and find a specific fluence of 0.2~Jy~ms for all three peaks.
FRB 181123's flux and multi-peak pulse profile resemble those from some previously discovered FRBs \citep{cpk+16}.
In particular, the two smaller peaks P2, P3, show narrow band features that resemble the down-drift pattern seen in the repeating bursts of FRB~121102 (\citealt{gsp+18, hessels19}; Li et al. in prep.). 

 \citet{hessels19} presented a detailed analysis of the complex time-frequency structures seen in the repeating bursts of FRB~121102. We followed their approach and estimated the drift rate between FRB~181123's P2 and P3 to be $\lesssim-140$~MHz~ms$^{-1}$ (Figure \ref{fig:dedisp}, right panel) in the observer's frame and $\lesssim-400$~MHz~ms$^{-1}$ in the rest frame of the FRB; the estimated drift rate has significant uncertainty, and it could be underestimated because we only see part of the spectrum of P2 and P3. 
Nevertheless, our estimated drift rate of $\lesssim-400$~MHz~ms$^{-1}$ fits well to the range measured in FRB 121102 \citet{hessels19}) around the rest frame emission frequency of 3--4.4~GHz, enhancing the similarities between FRB~181123 and FRB~121102.

\begin{figure}
\centering
\includegraphics[scale=0.7]{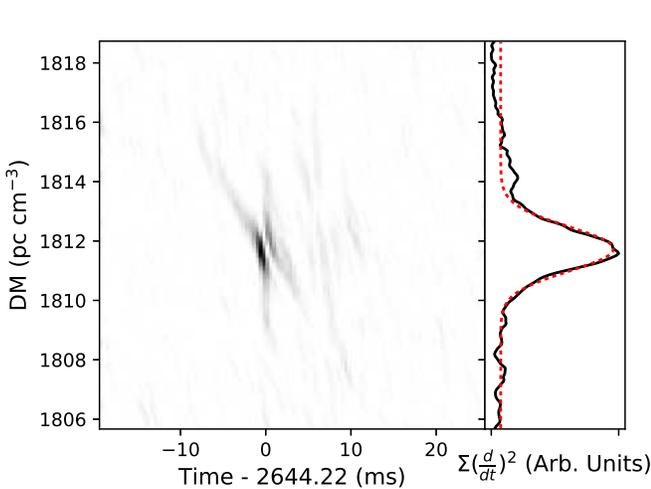} \\ 
\caption{\label{fig:DMcurve} 
{Similar to the Figure 2 of \citet{hessels19}, the left panel shows the
square of the Gaussian-smoothed forward difference time derivative of the
dedispersed burst profile as a function of DM and time. The profiles are
down sampled by a factor of 2 to boost the S/N. The right panel show the
sum along the time axis and its Gaussian fit. The bestfit DM from this
fine-structure analysis is $1812\pm1$~pc~cm$^{-3}$, consistent with the
estimate from the S/N of P1 (Table \ref{tab:par}).}
}
\end{figure}

Figure \ref{fig:DMcurve} shows the results of a fine frequency-time structure analysis applied to FRB~181123, following the approach in \citet{hessels19}. We found significant fine structures, characterized by the square of Gaussian-smoothed forward-difference time derivatives (i.e. the changes between every consecutive time bins), in the dedispersed bursts around the position of P1 and minor structures in P2 and P3. These fine structures allow us to derive the optimal DM as 1812$\pm$1~pc~cm$^{-3}$. This estimation is consistent with, and slightly more constrained than, what we we derived from using the S/N.
Unlike \citet{gsp+18} and \citet{hessels19}, we did not observe FRB~181123 in a coherent-dedispersion mode, thus the intra-channel smearing due to a DM of $\sim1812$ is 0.5~ms to 2~ms in our observing band. Hence,
DM smearing will not allow us to resolve structure finer than 0.5~ms despite our 0.196608~ms sampling interval.
The measured widths of P1, P2, and P3 are consistent with this DM-smearing width and show no significant evidence of scattering tails.

\begin{figure}
\centering
\includegraphics[scale=0.5]{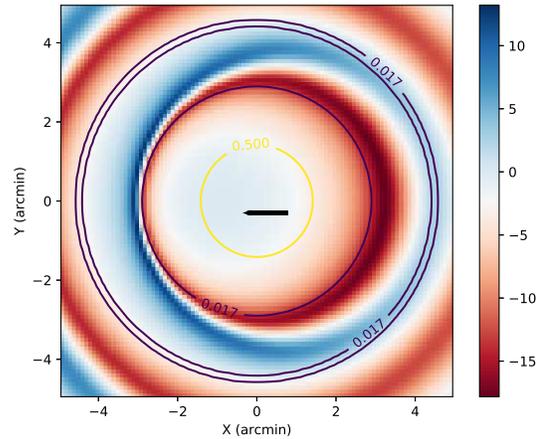} \\ 
\caption{\label{fig:beam} 
The circular contours show the theoretical power pattern of the FAST main beam at 1250~MHz, assuming a uniformly illuminated dish of 300~m diameter. The 0.5-level circle labels the half-power contour of the main beam. The two 0.017-level contours on the outside marks the rough position of the first sidelobe. The colored image underneath the contour shows the approximate extra spectral index caused by the frequency-dependent beam pattern and the source drifting. The black arrow illustrates how far a celestial object would drift in the dispersed duration of FRB~181123 (4.17~s). }
\end{figure}

As can be seen in the frequency structure plot in Figure ~\ref{fig:dedisp}, P1 of FRB~181123 is brighter at the lower frequency part of the band. From P1's on-pulse minus off-pulse spectrum shown in the middle panel of Figure ~\ref{fig:dedisp}, we find that the best-fit FRB spectrum index is $-3.3\pm0.5$. \citet{sch+14} detected the first burst of FRB~121102 in the sidelobe of the Arecibo beam. They argue that the sidelobe position varied with frequency and caused the detected burst spectrum to be steep and up-swinging (with spectral index 7--11). The same argument could be applied conversely to FRB~181123, in which case the FRB is likely detected by the main beam instead of the sidelobe. In contrast to the original observation of FRB~121102 \citep{sch+14}, FRB~181123 is detected in a drift scan where the beam was moving across the sky while the burst arrived, further changing the observed burst spectral index. 
We use a theoretical antenna power pattern to evaluate how these two factors change the FRB's spectral shape (Figure \ref{fig:beam}). 
The antenna power pattern of a uniformly illuminated dish,
\begin{equation}
    P(x,y,\nu)=[2J_1(u)/(u)]^2,
\end{equation} 
where 
\begin{equation}
u=\pi\sqrt{x^2+y^2} D \nu/c.
\end{equation} 
Here $x$ and $y$ represent the source position with respect to the beam center, $D$ is the dish diameter, $\nu$ is the observing frequency, $c$ is the speed of light and $J_1(u)$ is a Bessel function of the first kind \citep{ToRA}.
We integrate $P$ along the drifting path of the FRB 
\begin{equation}
I=\int G(\nu) (\nu/\nu_0)^2 P(x(t), y, \nu(t))dt, 
\end{equation}
to get an approximated power for two subbands: the bottom band (1000--1250~MHz) and the top band (1250--1500~MHz), here $G(\nu)$ is the gain of the telescope as a function of frequency $\nu$, and $(\nu/\nu_0)^2 $ is a normalizing factor with $\nu_0=1250$~MHz.
For convenience, we assume $G(\nu)$ to be flat while in practice it varies slightly with $\nu$ \citep{jth+20}.
The result of this integration depends on the assumed starting position of the FRB (i.e.~its position at 1500~MHz), and the FRB's DM value.
We then use the ratio between the integrated power in the up and bottom band to derive an approximation for the extra spectral index: $\Delta \gamma \sim \log(I_{\rm top}/I_{\rm bottom})/\log(\rm \nu_{\rm top}/\nu_{\rm bottom})$, where $\nu_{\rm top} = 1375$~MHz and $\nu_{\rm bottom}=1125$~MHz.
As shown in Figure ~\ref{fig:beam}, if FRB~181123 were detected in the sidelobe, its spectrum would likely have been significantly impacted. But, we observed a relatively flat burst spectrum, and the main peak P1's signal persists across the whole band. This suggests that FAST likely caught the FRB in the main lobe.
Admittedly, the observed antenna pattern of FAST \citep{jth+20} may be quantitatively different from a theoretical one, but our conclusion should still be valid.

The observed dispersion measure of FRB~181123 ($1812\pm 2$~pc~cm$^{-3}$) includes  contributions from the intergalactic medium (IGM) -- DM$_{\rm IGM}$, from the Galaxy -- DM$_{\rm Gal}$ and from the host galaxy of the FRB -- DM$_{\rm host}$. We assume DM$_{\rm Gal}\sim149.5$ based on \citet{ymw16} and DM$_{\rm host}\sim 40/(1+z)$~pc~cm$^{-3}$ \citep{xh15,yz16}, where $z$ is the redshift of the host galaxy. 
\citet{zhang18} derived how one could estimate the upper limit of FRB luminosity based on its observed DM. They also derived a DM--$z$ relation that correctly accounts for the integrated dispersion effect for objects in cosmological distances assuming homogeneous IGM, and provided an approximated formula $z\sim$DM$_{\rm IGM} / 855$~pc~cm$^{-3}$ for $z<2$.
We follow their calculations closely and solve for the redshift $z$ of FRB 181123 using the equation ${\rm DM_{obs}-DM_{Gal}} = (855z + 40/(1+z))$~pc~cm$^{-3}$.
The best solution is $z\lesssim $1.93 and DM$_{\rm IGM}\simeq1650$~pc~cm$^{-3}$. 
Note that we kept 3 significant digits for $z$ and DM$_{\rm IGM}$ to show the exact solution to the above equation.

To quantify the uncertainties on the
above $z$ estimate, we now step through the relevant contributions.
We note that the estimated DM$_{\rm Gal}$ has $\sim$50\% uncertainty \citep{ymw16}, DM$_{\rm host}$ may contain $\sim $100\% uncertainty, together they contribute 5\% relative uncertainty to the estimated DM$_{\rm IGM}$.
Furthermore, due to inhomogeneity in the IGM, objects from the same DM$_{\rm IGM}$ could be from different $z$ \citep{plm+19}, according to \citet{wmb18}, this could increase the uncertainty in our estimated $z$ by an additional factor of 10\%.
In the above derivations, we assumed probable distributions of DM$_{\rm Gal}$ and DM$_{\rm host}$, but we could still underestimate them substantially and thus overestimate DM$_{\rm IGM}$ and $z$. 
Therefore, we treat the FRB's derived redshift, luminosity, and energy as upper limits.

Assuming the $\Lambda$CDM cosmological parameters with $H_0=67.8\pm0.9$~km~s$^{-1}$~kpc 
and $\Omega_{\rm M} =0.308\pm0.012$ \citep{planck16}, the luminosity distance of a $z=1.93$ object is $\simeq15.3$~Gpc \footnote{\url{https://docs.astropy.org/en/stable/cosmology}}.
Based on equation (8) and (9) in \citet{zhang18}, FRB~181123's peak luminosity is $\gtrsim 2\times10^{43}$~erg~s$^{-1}$ and the isotropic energy $\gtrsim 2\times10^{40}$~erg, both limits contain relative uncertainties of $\sim 15$\%.
These values are comparable to those derived in Table 1 of \citet{zhang18} from previously discovered FRBs.

\begin{deluxetable}{lc}
\tablecaption{ \label{tab:par}
Observational Parameters of FRB 181123.}
\tabletypesize{\scriptsize}
    \tablewidth{0pt}
    \tablehead{\colhead{Parameter}	&	\colhead{Value
    \tablenotemark{a}	}}
\startdata
Date (UTC) & 2018 Nov 23\\
Time (UTC) & 17:49:09  \\
MJD arrival time\tablenotemark{b}	& 58445.74246675\\
Right ascension\tablenotemark{c}	&	05$^{\rm h}$06$^{\rm m}$06$^{\rm s} $.76\\
Declination\tablenotemark{c}		&	18$^{\degr}$09$^{'}$35$^{''} $.7 \\
Gal. long.	&	184.06\\
Gal. lat.	&	--13.47\\
DM (pc~cm$^{-3}$)	&	1812(2) \\
P1 pulse arrival time\tablenotemark{d}(s)	&	2.64422(4)\\
P1 pulse height (mJy) &	65(3)\\
P1 pulse width (ms) &	 1.05(6)\\
P2 pulse arrival time\tablenotemark{d} (s)	&   2.6499(1)	\\
P2 pulse height (mJy) &	24(3)\\
P2 pulse width (ms) & 0.6(2)	\\
P3 pulse arrival time\tablenotemark{d} (s)	&2.6538(1)	\\
P3 pulse height (mJy) &19(2)	\\
P3 pulse width (ms) &1.2(2)	\\
P1 Spectral index\tablenotemark{e} &   --3.3(5) \\
P2-P3 frequency drift rate\tablenotemark{f} & $\lesssim$--140~MHz~ms$^{-1}$ 
\enddata
    \tablenotetext{a}{Number in parentheses indicates 1-$\sigma$ uncertainty in the least significant digit.}
    \tablenotetext{b}{The topocentric arrival time of the first peak (P1) in MJD.}
    \tablenotetext{c}{Coordinates obtained from the position of the
receiver cabin when P1 arrived. We assume a position error circle of 3' radius.}
    \tablenotetext{d}{Pulse arrival time at the top edge of the band
(1500~MHz) since the start of the particular data file.}
    \tablenotetext{e}{$S(\nu)\propto\nu^\alpha$}
    \tablenotetext{f}{Measured in the observer's frame.}
\end{deluxetable}

\subsection{A lower bound on the FAST event rate}
Since the PICS-aided GPU searching system is an experimental pipeline, it probably does not find all events that cross the 7$\sigma$ threshold. A more thorough search is currently being conducted using standard software such as {\it HEIMDALL}.
With one detection from a pipeline of recall $<1$, we can only calculate a lower bound on the FAST event rate for the given detection threshold of 7$\sigma$, i.e., 25~mJy~ms for 1-bit polarization-summed data. 
Assuming that the FRB events follow a Poisson distribution, the probability density distribution of the first detected event should follow an exponential distribution, i.e., Poisson distribution of zero events until the first detection. 
In this case, the cumulative distribution function of an exponential
distribution follows $F(x) =1 - e^{-k x} $, where $k$ is the
event rate, and $x$ is the time to the first detection. 
Here we would like to find the 95\% confidence limits for the event rate
$k$ giving one event in 1500 hours. We find that $F(1500\ \rm hours) > 0.05$ when $k > 0.034$ event per 1000 hour (i.e. 0.3 event per year).
Considering that our search system does not have a 100\% recall rate, no upper bound can be set. 
Due to small number statistics, the lower limit of $k > 0.034 $ event per 1000 hour is not yet constraining to most theoretical predictions \citep{lhz+17, lllz18, lml+20}. 

Nevertheless, this first detection attests to FAST's potential to systematically detect FRBs in the future, and such detections will put far more stringent constraints on the FRB-rate at high DM.
\citet{lor18} and \citet{zhang18} showed that the FAST event rate, especially the rate of high-DM FRBs, would help determine the luminosity function of FRBs.
Detection of very high DM ($> 6500$ pc~cm$^{-3}$) could probe FRB at more than $z=10$ \citep{zhang18}, and help shed light on the cosmological distribution of the FRBs.

\section{Discussion and Summary}

FRB~181123 shows a clear multi-peak profile. Its two smaller peaks, P2 ($5.7\pm0.2$~ms from P1), and P3 ($3.9\pm0.2$~ms from P2) show narrow band features that resemble the down-drifting pattern seen in the bursts of repeating FRB~121102 \citep{gsp+18,hessels19}, 181128, 181222 and 181226 \citep{chime_8}. 
Although multiple sub-bursts and fine pulse structures have also been observed from (so far) non-repeating FRBs \citep{cpk+16,ffb+18,cms+20}, the combination of multiple sub-bursts (or components) with millisecond spacing
and down-drifting pattern have mostly been seen from repeating FRBs. This suggests that FRB~181123 could be a repeating FRB source. To test this, we conducted follow-up observations toward the position of the FRB using FAST. So far, we observed the position during for four independent sessions each with one-hour integration on 2020 February 2, 28, and 29 and 2020 March 27. We have not detected any repeating bursts above the fluence level 0.012~Jy~ms (7$\sigma$ limit; we used 8-bit digitization and two polarization in the follow-up observations, thus reached a lower detection threshold than in the original 1-bit data).  
The non-detection of repeating bursts from FRB~181123 may be due to one of the following four reasons \citep[e.g.][]{palaniswamy18}: 1. The waiting times for producing repeating bursts may be longer than the duration of our follow up; 2. Faint repeating bursts may be produced in our observing window, but are below the detection threshold. This requires that the peak fluxes of the putative repeating bursts are lower than that of FRB~181123 by a factor of more than six\footnote{A similar case has been observed in FRB 171019 \citep{kumar19}, whose repeating bursts are much fainter than the originally detected burst.}; 3. The burst activity may be intermittent (e.g. changing due to unidentified periodic activity) like in FRB 121102, and our follow up observations may be taken when the source is not active; 4. The source is an intrinsically non-repeating FRB. It is possible that either of the first three reasons are at play. We plan to continue monitoring the FRB in the coming months and hopefully will eventually detect some bursts if the FRB is a repeater.

The last possibility is difficult to prove, but if true, a catastrophic event has to be able to produce multiple peaks during the emission process. This is challenging for most models, even though in some scenarios this may be possible. For example, the ``blitzar'' model \citep{fr14} suggests that an FRB could originate from the final flash of a supermassive neutron star collapsing into a black hole by magnetic braking. Detailed simulations \citep{most18} showed that this scenario can produce a series of sub-ms pulses whose amplitudes decay exponentially with time. The observed duration of the sub-pulses of FRB 181123, when corrected for the redshift factor, may be consistent with this model. The down-drifting feature seen in sub-pulses may be understood within the generic bunching curvature radiation model invoking open field lines \citep{wang19}, which is invoked in the blitzar model during the magnetospheric ejection phase \citep{most18}.

FAST's sensitivity makes it one of the most effective telescope at detecting FRBs from high redshift, therefore its FRB detection rate is an important observable. 
\citet{lhz+17} predicted that the FRB detection rate for the FAST 19-beam would be $5\pm2$ detections per 1000 hours, based on an all-sky event rate of $3\times10^4$ day$^{-1}$ that crosses an energy threshold of 0.03~Jy~ms.
From a different approach, by measuring the event rate density of the luminosity function presented in \citet{lllz18}, \citet{lml+20} predicted an all-sky event rate of $10^4$--$10^5$~day$^{-1}$ for events with flux higher than 5~mJy, which correspond to 1.5--15 events per 1000 hours given the field of view of the FAST 19-beam.
With one detection of FRB 181123, we 
can place a lower bound of $0.034$ event per 1000 hour,
which can be translated to an all-sky rate of $>9\times10^2$ day$^{-1}$.

If FRB~181123 is a one-off FRB, not a repeater, it may be possible that at least some energetic FRBs may form a distinct category from the repeaters. 
Then one may use this event to estimate the event rate density of these energetic events and compare it with some models predicting one-off FRBs, e.g. those models invoking compact star mergers. 
Equation (10) in \citet{zhang18} shows how the fluence ($F_{\nu}\simeq  S_{\nu}\tau_{\rm obs}$) of a putative FRB scales with redshift: \begin{equation}
    F_{z'} = \left(\frac{1+z'}{1+z}\right)^{1+\alpha}\left(\frac{D_{\rm L}}{D_{\rm L}'}\right)^2F_{z},
\end{equation} 
where $D_{\rm L}$ and $D_{\rm L}'$ denote the luminosity distances corresponding to redshift $z$ and $z'$, and $\alpha$ is the spectra index of the FRB.
Following this equation, we find that a FRB like FRB~181123 could be detected with
0.025~Jy~ms fluence at a redshift of $z\simeq4.25$. 
For the value of $\alpha$ we used the observed spectral index of P1.
The resulting redshift corresponds to a comoving volume of 1800~Gpc$^3$. 
So far, we have made a single detection of an FRB event above $10^{40}$~erg in energy in a volume of 1800~Gpc$^3$. 
For the total amount of data we searched, we could infer the 95\% confidence lower limit event rate of 900 per day, and an event rate density lower limit of $>200$~Gpc$^{-3}$yr$^{-1}$ for FRBs with energy $>10^{40}$~erg. 
This lower limit is underestimated because we can only observe a fraction of the FRBs, some (maybe most) FRB events have a lower isotropic energy than $10^{40}$~erg, some FRBs may be beamed and likely not beaming towards us. 
Interestingly, this lower limit is already in mild tension with the black hole-black hole (BH-BH) merger event rate density ($\sim 200$~Gpc$^{-3}$yr$^{-1}$) inferred from LIGO observations \citep{mg18} (regardless whether BH-BH mergers can make FRBs), but could be consistent with neutron star - neutron star (NS-NS) merger event rate density ($\sim 1.5\times10^3$~Gpc$^{-3}$~yr$^{-1}$; \citealt{aaa+17}). More detections may be made in the same dataset we used and the true event rate density may be better constrained to a (much) higher value than our limit. This could lead to better constraints on the event rate density of energetic events \citep{lml+20}, giving tighter constraints on the consistency with the compact star merger models \citep[see also][]{wang20}. 
FAST is a very sensitive telescope. The FRB sample from FAST blind search
would likely be composed of many high DM, high redshift events with higher
isotropic energy than samples from other telescopes. Therefore, the FAST blind
search FRB sample may become relevant for testing catastrophic models for ``one-off'' FRBs.

\acknowledgements
The authors thank Shu-Xu Yi, Nan Li, and Zhi-Yuan Ren for discussions and the referee for a careful review and suggestions.
This work is supported by National Key R\&D Program of China No.
2017YFA0402600, the CAS-MPG LEGACY project and the FAST FRB key science project. 
WWZ is supported by the CAS Pioneer Hundred Talents
Program, the Strategic Priority Research Program of the CAS Grant No. XDB23000000, and by the National Natural Science Foundation
of China under grant No. 11690024, 11743002, 11873067. 
LQ is supported in part by the Youth Innovation Promotion Association of CAS
(id.~2018075).
YLY is supported by CAS "Light of West China" Program.
ZCP is supported by the National Natural Science Funds of China (Grant No.
11703047) and the CAS "Light of West China" Program.
DRL is supported by National Science Foundation OIA Award 1458952.
This research made use of Astropy,\footnote{http://www.astropy.org} a community-developed core Python package for Astronomy \citep{astropy:2013, astropy:2018}. 
This work is supported by Chinese Virtual Observatory (China-VO) and Astronomical Big Data Joint Research Center, co-founded by National Astronomical Observatories, Chinese Academy of Sciences and Alibaba Cloud.
{\it Facilities:}
\facility{Five-hundred-meter  Aperture  Spherical  radio  Telescope  (FAST)}


\bibliographystyle{apj.bst}
\bibliography{myrefs}

\end{document}